\documentclass[aps,pra,twocolumn,showpacs,superscriptaddress,floatfix]{revtex4}

\usepackage{graphicx}
\usepackage{nicefrac}
\usepackage{amsmath}
\usepackage{amsfonts}
\usepackage{amssymb}
\usepackage{amsthm}
\usepackage{epsf}
\usepackage{bm}
\usepackage{bbm}
\usepackage{longtable}

\usepackage{dcolumn}

\def\Za{{Z\alpha}}
\def\pr{^{\prime}}
\def\rms{\left<r^2\right>^{1/2}}
\def\eps{\epsilon}

\newcommand{\cross}[1]{#1\!\!\!/}

\newcolumntype{.}{D{x}{}{-1}}

\newcommand{\vare}{\varepsilon}
\newcommand{\bfx}{{\bm {x}}}

\newcommand{\bfp}{{\bm {p}}}
\newcommand{\bfq}{{\bm {q}}}

\newcommand{\lbr}{\langle}
\newcommand{\rbr}{\rangle}

\newcommand{\intinf}{\int^{\infty}_{-\infty}}

\newcommand{\balpha}{\bm{\alpha}}

\begin{document}

\title{Two-loop QED corrections with closed fermion loops}

\author{V.~A.~Yerokhin}
 \affiliation{Center for Advanced Studies, St.~Petersburg State Polytechnical
University, Polytekhnicheskaya 29, St.~Petersburg 195251, Russia}

\author{P.~Indelicato}
\affiliation{Laboratoire Kastler-Brossel, \'Ecole Normale Sup\'erieure, CNRS et
Universit\'e P. et M. Curie, Case 74, 4 pl.~Jussieu, F-75252, France}

\author{V.~M.~Shabaev}
\affiliation{Department of Physics, St.~Petersburg State University, Oulianovskaya
1, St.~Petersburg 198504, Russia}

\begin{abstract}

We report a calculation of all two-loop QED corrections with closed fermion loops
for the $n=1$ and $n=2$ states of H-like ions and for a wide range of the nuclear
charge numbers $Z=1-100$. The calculation is performed to all orders in the
binding-strength parameter $\Za$, with the exception that in a few cases the
free-loop approximation is employed in the treatment of the fermion loops.
Detailed comparison is made with previous $\Za$-expansion calculations and the
higher-order remainder term to order $\alpha^2(\Za)^6$ is identified.

\end{abstract}

\pacs{31.30.jf, 31.15.A-, 31.10.+z}

\maketitle

\section*{Introduction}

Highly charged ions are often considered as the ideal testing ground for
investigating the strong-field regime of bound-state quantum electrodynamics
(QED). They feature a strong static Coulomb field of the nucleus and have a simple
electronic structure, which can be accurately determined in {\em ab initio}
theoretical calculations.  In this respect, the ultimate investigation object is
the H-like uranium, the heaviest naturally occurring element. Measurements of the
$1s$ Lamb shift in H-like uranium have progressed drastically during last decades
\cite{stoehlker:93,stoehlker:00:prl}, having achieved an accuracy of 4.6~eV
\cite{gumberidze:05:prl}, which corresponds to a fractional error of 1.7\% with
respect to the total QED contribution. Further advance anticipated in the future
will make such experiments sensitive to the two-loop QED effects.

Even higher precision is achieved for heavy Li-like ions. Measurements of the
$2p_{1/2,3/2}$-$2s$ transition energies
\cite{schweppe:91,beiersdorfer:98,bosselmann:99,brandau:04,beiersdorfer:05} have
lately reached a fractional accuracy of 0.03\% with respect to the total QED
contribution. This corresponds to a 10\% sensitivity of the experimental results
to the two-loop QED effects. These measurements provide an excellent possibility
for the identification of the two-loop Lamb shift and for testing the bound-state
QED up to second order in $\alpha$ in the strong-field regime ($\alpha$ is the
fine-structure constant). Theoretical description of Li-like ions is more
complicated than that of H-like ions, which is due to the presence of additional
electrons. For heavy systems, the electron-electron interaction is weak (as
compared to the electron-nucleus interaction) and can be successfully treated by
perturbation theory. {\em Ab initio} calculational results for the effects of the
electron correlation and the screening of one-loop QED corrections are already
available for Li-like ions \cite{yerokhin:00:prl,sapirstein:01:lamb}; their
accuracy is sufficient for identification of the two-loop QED effects.

The detailed theoretical understanding of the two-loop QED effects is also
necessary for the interpretation of high-precision experimental data in the
low-$Z$ region. The most prominent example here is hydrogen. Its spectroscopy can
nowadays be realized with a relative accuracy on the level of $10^{-14}$
\cite{niering:00,fischer:04}. The theoretical understanding of the $1s$ and $2s$
Lamb shift in hydrogen is still limited, to a large extent, by the two-loop QED
effects \cite{mohr:05:rmp}.

\begin{figure}[b]
\centerline{
\resizebox{0.95\columnwidth}{!}{%
  \includegraphics{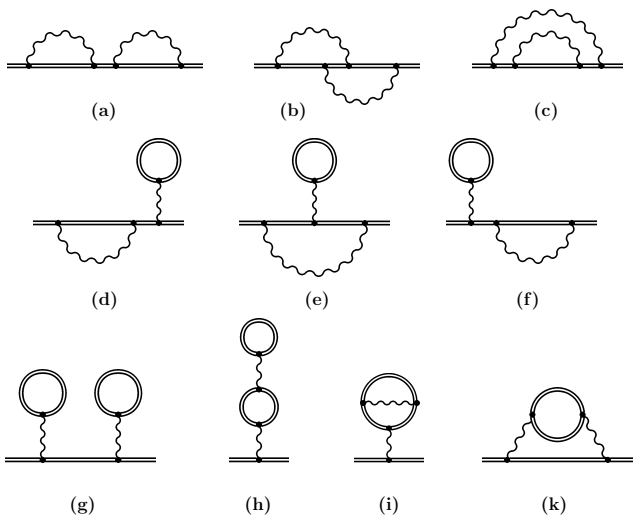}
}}
 \caption{Two-loop one-electron QED corrections. Gauge-invariant subsets are
referred to as SESE (a-c), SEVP (d-f), VPVP (g-i), and S(VP)E (k).
\label{fig:2order}}
\end{figure}

The subject of our present interest is the set of the two-loop QED corrections
(also referred to as the two-loop Lamb shift), graphically represented in
Fig.~\ref{fig:2order}. These corrections have been intensively investigated within
the perturbative expansion in the binding-strength parameter $\Za$
\cite{pachucki:01:pra,jentschura:02:jpa,pachucki:03:prl,czarnecki:05:prl} ($Z$ is
the nuclear charge number). The results of these studies, however, do not provide
reliable information about the magnitude of the two-loop Lamb shift in heavy ions,
where the parameter $\Za$ approaches unity. A non-perturbative (in $\Za$)
evaluation of the whole set of the two-loop diagrams is needed for the
interpretation of experimental results available in the middle- and high-$Z$
region.

Numerical all-order calculations of the two-loop corrections started in late 1980s
\cite{beier:88,schneider:93:ks,persson:96:pra,mallampalli:96,beier:97:jpb,plunien:98:epj}
and dealt with the diagrams with the closed fermion loops
[Fig.~\ref{fig:2order}(d)-(k)]. An evaluation of the three remaining diagrams
[Fig.~\ref{fig:2order}(a)-(c)], referred to as the two-loop self-energy diagrams,
turned out to require considerable efforts. It was accomplished in a series of
investigations
\cite{mitrushenkov:95,mallampalli:98:pra,yerokhin:01:sese,yerokhin:03:prl,%
yerokhin:03:epjd,yerokhin:05:sese,yerokhin:05:jetp,yerokhin:06:prl} for the
nuclear charge numbers $Z\geq 10$ for the ground state and $Z\geq 60$ for the
$n=2$ states.

The goal of this work is to perform a detailed investigation of all two-loop
diagrams with the closed fermion loops [Fig.~\ref{fig:2order}(d)-(k)], extending
previous evaluations to cover the whole region of the nuclear charge numbers
$Z=1-100$ and all $n=1$ and $n=2$ states. The first results of our calculation
were presented in Ref.~\cite{yerokhin:06:prl}. At present, our intention is to
achieve high numerical accuracy in the low-$Z$ region. This will allow us to make
a detailed comparison with the perturbative calculations and to isolate the
non-perturbative remainder to order $\alpha^2(\Za)^6$, which is of experimental
interest for hydrogen. Our calculation will be performed to all orders in the
binding-strength parameter $\Za$, but an approximation will be made in the
treatment of the fermion loops in the diagrams in Fig.~\ref{fig:2order}(h)-(k).
This approximation, referred to as the {\em free-loop} one, consists in keeping
the leading nonvanishing term of the expansion of the fermion loop in terms of the
binding potential. In the one-loop case, it corresponds to the Uehling potential
and yields the dominant contribution even for high-$Z$ ions. To perform a
non-perturbative calculation of the diagrams in Fig.~\ref{fig:2order}(h)-(k)
without the free-loop approximation is a difficult and a so far unsolved problem.

The paper is organized as follows. In Section~\ref{sec:oneloop}, we describe our
approach to the evaluation of the one-loop self-energy and vacuum-polarization
corrections, which is used as a basis for treatment of the two-loop corrections.
In Section~\ref{sec:twoloop}, we discuss the evaluation of individual two-loop
corrections with closed fermion loops, present our numerical results, and compare
them with the data obtained within the perturbative $\Za$-expansion approach. In
Section~\ref{sec:concl}, we summarize our results.

The relativistic units ($\hbar=c=1$) are used in this paper.

\section{One-loop QED corrections} \label{sec:oneloop}

We start with presenting some basic formulas for the first-order self-energy and
vacuum-polarization corrections that will be needed below in addressing the
two-loop corrections.

\subsection{Self-energy} \label{sec:se}

The one-loop self-energy (SE) correction to the Lamb shift can be represented as a
matrix element of the renormalized SE operator $\Sigma_R$,
\begin{equation}\label{i5}
\Delta E_{\rm SE} = \lbr  a|\gamma^0
    \Sigma_R(\vare_a) |a\rbr\,,
\end{equation}
where $\Sigma_R = \Sigma -\delta m$, $\delta m$ is the one-loop mass
counterterm, $\vare_a$ is the energy of the reference state, $\Sigma$ is the
unrenormalized SE operator,
\begin{eqnarray} \label{i6}
 \Sigma(\vare,\bfx_1,\bfx_2) &=&
2i \alpha\gamma^0 \intinf d\omega\,
      \alpha_{\mu}\,
 \nonumber \\ && \times
         G(\vare-\omega,\bfx_1,\bfx_2)\, \alpha_{\nu}\,
    D^{\mu\nu}(\omega,\bfx_{12})\,
         \,,
\end{eqnarray}
$G$ is the Dirac Coulomb Green function $G(\vare) = [\vare-{\cal H}(1-i 0)]^{-1}$,
$\cal H$ is the Dirac Coulomb Hamiltonian, $D^{\mu\nu}$ is the photon propagator,
${\alpha}^{\mu} = (1, \balpha)$, and $\bfx_{12} = \bfx_1-\bfx_2$. It is assumed
that the above expressions are regularized in a covariant way and that the limit
that removes the regularization is properly approached.

In order to facilitate the numerical evaluation of the above expressions, one
needs to explicitly eliminate divergences from the SE operator. A popular method
of doing this \cite{snyderman:91} is based on the expansion of the SE operator in
terms the binding field,
\begin{equation}\label{i7}
\Sigma = \Sigma^{(0)} + \Sigma^{(1)}+ \Sigma^{(2+)}\,,
\end{equation}
where the superscript $(i)$, $i = 0,1$ indicates the total number of interactions
with the binding Coulomb field and the index $(2+)$ labels the term generated by
$\geq 2$ such interactions. Only the first two terms of the expansion are
divergent; all divergences can be shown to vanish in the difference $\Sigma
-\delta m$. The divergent parts are regularized by working in an extended number
of dimensions and evaluated in momentum space, see Ref.~\cite{yerokhin:99:pra} for
details. We mention that the first-order expansion term $\Sigma^{(1)}$ is usually
represented as a product of the time component of the standard vertex operator
$\Gamma^{\alpha}(p_1,p_2)$ and the Coulomb potential $V_C$,
\begin{equation}\label{i7a}
\Sigma^{(1)}(p_1,p_2) \equiv \Gamma^0(p_1,p_2)\,V_C(|\bm{p_1}-\bm{p_2}|)\,.
\end{equation}

The energy shifts induced by (the final parts of) the three terms in the
right-hand-side of Eq.~(\ref{i7}) are referred to as the zero-, one-, and
many-potential terms, respectively:
\begin{equation}\label{i8}
\Delta E_{\rm SE} = \Delta E^{(0)}_{\rm SE}+
        \Delta E^{(1)}_{\rm SE}+ \Delta E^{(2+)}_{\rm SE} \ ,
\end{equation}
where
\begin{align}            \label{i9}
  \Delta E^{(0)}_{\rm SE}
= \int \frac{d\bfp}{(2\pi)^3} \,
        \overline{\psi}_a(\bfp)\,  \Sigma^{(0)}_R(p)\, \psi_a(\bfp) \, ,
\end{align}
\begin{align}                \label{i10}
 \Delta E^{(1)}_{\rm SE} =
        \int \frac{d\bfp_1 d\bfp_2}{(2\pi)^6}\,
        \overline{\psi}_a(\bfp_1)\,
        \Gamma^0_R(p_1,p_2)\,
        V_C(\bfq)\, \psi_a(\bfp_2) \, ,
\end{align}
\begin{align}   \label{i10a}
  \Delta E^{(2+)}_{\rm SE}
= \int d\bfx_1 d\bfx_2\,
        \overline{\psi}_a(\bfx_1)\,  \Sigma^{(2+)}(\vare_a,\bfx_1,\bfx_2)\, \psi_a(\bfx_2) \, ,
\end{align}
where $\bfq = \bfp_1-\bfp_2$, $\overline{\psi}_a = \psi_a^{\dag}\,\gamma^0$,
$V_C(\bfq) = -4\pi\Za/|\bfq|^2$ is the Coulomb potential in momentum space,
$\Sigma^{(0)}_R(p)$ is the finite part of the subtracted free SE operator
$\Sigma^{(0)}(p)-\delta m$, $\Gamma^{\mu}_R(p_1,p_2)$ is the finite part of the
vertex operator $\Gamma^{\mu}(p_1,p_2)$; and $p$, $p_1$, and $p_2$ are
four-vectors with the time component fixed by $p^0 = p_1^0 = p_2^0 = \vare_a$. The
operator $\Sigma^{(2+)}$ is obtained from Eq.~(\ref{i6}) by the substitution $G
\to G^{(2+)}$, where the index $(2+)$ denotes, as usual, two or more interactions
with the binding Coulomb field.

In the extended number of dimensions, $D = 4-2\eps$, the free SE
operator $\Sigma^{(0)}$ is given by
\begin{equation}\label{i11}
  \Sigma^{(0)}(p) = -4\pi i \alpha \mu^{2\eps} \int \frac{d^D
  k}{(2\pi)^D}\,\frac1{k^2}\,
     \frac{\gamma_{\sigma}(\cross{p}-\cross{k}+m)\gamma^{\sigma}}{(p-k)^2-m^2}\,,
\end{equation}
where $\cross{p} = p^{\nu}\gamma_{\nu}$ and $\mu$ is the auxiliary mass parameter
introduced in order to keep the proper dimension of the interaction term in the
Lagrangian. The momentum integration in Eq.~(\ref{i11}) is simple; it is
described, e.g., in Appendix A of Ref.~\cite{yerokhin:03:epjd}. Omitting terms of
order $\eps$ and higher, one obtains
\begin{eqnarray} \label{i11a}
\Sigma^{(0)}(p)-\delta m &=& -\frac{\alpha C_{\eps}}{4\pi\eps}(\cross{p}-m)
      \nonumber \\ &&
  +\frac{\alpha}{4\pi}
\Biggl[ 2m \left( 1 +\frac{2\rho}{1-\rho}
    \ln \rho \right)
      \nonumber \\ &&
 - \cross{p}\, \frac{2-\rho}{1-\rho} \left( 1+\frac{\rho}{1-\rho} \ln \rho \right)
        \Biggr]\,,
\end{eqnarray}
where $\rho = (m^2-p^2)/m^2$ and $C_{\eps} = \Gamma(1+\eps)(4\pi)^{\eps}
\left(\mu^2/m^2\right)^{\eps}$. The operator $\Sigma^{(0)}_R$ in Eq.~(\ref{i9}) is
the finite part of the right-hand-side of Eq.~(\ref{i11a}) when $\eps$ approaches
zero.

The free vertex operator $\Gamma^{\alpha}$ is
\begin{eqnarray}\label{i12}
\Gamma^{\alpha}(p_1,p_2) &=& -4\pi i \alpha \mu^{2\eps} \int
    \frac{d^Dk}{(2\pi)^D} \frac1{k^2} \gamma_{\sigma}
    \frac{\cross{p}_1-\cross{k}+m}{(p_1-k)^2-m^2}
\nonumber \\ && \times
     \gamma^{\alpha}
        \frac{\cross{p}_2-\cross{k}+m}{(p_2-k)^2-m^2} \gamma^{\sigma} \, .
\end{eqnarray}
The momentum integration is performed by using the standard technique (see, e.g.,
Appendix A of Ref.~\cite{yerokhin:03:epjd}). Omitting terms of order $\eps$ and
higher, one obtains
\begin{eqnarray} \label{i12a}
\Gamma^{\alpha}(p_1,p_2) &=&
 \frac{\alpha C_{\eps}}{4\pi \eps}\,\gamma^{\alpha}
 -\frac{\alpha}{4\pi}
   \Biggl[\frac32 \gamma^{\alpha}
\nonumber \\ &&
 +\int_0^1 dx dy\,
    \left(
     \frac{N^{\alpha}(xb)}{D}+ 2x \gamma^{\alpha} \ln D \right)
   \Biggr] \,,
\end{eqnarray}
where $N^{\alpha}(xb) = \gamma_{\sigma}(\cross{p}_1-x\cross{b}+m)\gamma^{\alpha}
(\cross{p}_2-x\cross{b}+m)\gamma^{\sigma}$, $b = yp_1+(1-y)p_2$, and $D = x b^2
+m^2 -y p_1^2 -(1-y) p_2^2$. The operator $\Gamma^{\alpha}_R$ in Eq.~(\ref{i9}) is
the  finite part of the right-hand-side of Eq.~(\ref{i12a}) when $\eps$ approaches
zero.

Out of the three terms in Eq.~(\ref{i8}), the third one [$\Delta E^{(2+)}_{\rm
SE}$] is the most difficult to evaluate numerically. To a large extent, this is
due to the partial-wave expansion that inevitably appears in the evaluation of the
function $G^{(2+)}$. The convergence of this expansion is rather good for the
ground state of high-$Z$ ions (provided that radial integrations are carried out
first), but quickly worsens when $Z$ is decreased and(or) the principal quantum
number $n$ is increased. An approach to overcome this difficulty was suggested in
Ref.~\cite{yerokhin:05:se}. It is based on the separation of the $\geq2$-potential
Green function $G^{(2+)}$ into two parts,
\begin{align}\label{i13}
    G^{(2+)}(\vare,\bfx_1,\bfx_2)&\  = G^{(2+)}_a(\vare,\bfx_1,\bfx_2)
  \nonumber \\ &
    +
     \bigl[G^{(2+)}(\vare,\bfx_1,\bfx_2)-G^{(2+)}_a(\vare,\bfx_1,\bfx_2)\bigr]\,,
\end{align}
where $G^{(2+)}_a$ yields an approximation to $G^{(2+)}$ in the region where
$\bfx_1 \approx \bfx_2$ and the energy argument is far away from a pole. This
separation is related to the one used  in Ref.~\cite{indelicato:92:pra} for the
identification of ultraviolet divergent parts of the SE operator in coordinate
space. The function $G_a^{(2+)}$ is given by
\begin{align}\label{i13a}
    G_a^{(2+)}(\vare,\bfx_1,\bfx_2)  =&\
G^{(0)}(\vare+\Omega,\bfx_1,\bfx_2)
   - G^{(0)}(\vare,\bfx_1,\bfx_2)
           \nonumber \\ &
   -         \Omega\, \left.\frac{\partial}{\partial E}\, G^{(0)}(E,\bfx_1,\bfx_2)
   \right|_{E=\vare}\,,
\end{align}
where $G^{(0)}$ is the free Dirac Green function and $\Omega  = 2\Za/(x_1+x_2)$.
The function $G^{(2+)}_a$ has two important properties: (i) it can be easily
expressed both in the closed form and in the partial-wave expansion form and (ii)
it depends on angular variables through $\bfx_{12} = \bfx_1-\bfx_2$ only. These
features ensure that the numerical evaluation of expressions with $G^{(2+)}_a$ is
rather simple. The main reason to introduce the auxiliary function $G^{(2+)}_a$ is
that the separation (\ref{i13}) improves the convergence of the partial-wave
expansion. It was demonstrated in Ref.~\cite{yerokhin:05:se} that the substitution
of the difference $G^{(2+)}-G^{(2+)}_a$ into (the high-energy part of)
Eq.~(\ref{i10a}) improves the convergence of the partial-wave expansion
drastically and allows one to obtain reasonably accurate results for the SE
correction at $Z=1$ with employing just about 30 partial waves. This approach was
extensively used in the present work for the evaluation of the two-loop
corrections. Its usage allowed us to obtain accurate numerical results for the
whole region of the nuclear charge numbers $Z=1-100$ in calculations of the
diagrams in Fig.~\ref{fig:2order}(d)-(f) and (k).

\subsection{Vacuum-polarization}
\label{sec:vp}

 The one-loop vacuum-polarization (VP) potential consists of two parts, which
are commonly referred to as the Uehling and the Wichmann-Kroll ones, see the
review \cite{mohr:98} for details,
\begin{equation}\label{ii1}
    U_{\rm VP}(r) = U_{\rm Ueh}(r) + U_{\rm WK}(r)\,.
\end{equation}
The expression for the Uehling potential is given by
\begin{align}\label{c2}
   U_{\rm Ueh}(r) &\ = -\frac{2\alpha^2 Z}{3 m r}\,
    \int_0^{\infty} dr\pr r\pr \rho(r\pr)\,
  \nonumber \\ & \times
       \left[ K_0(2m|r-r\pr|)-K_0(2m|r+r\pr|)\right]\,,
\end{align}
where
\begin{equation}\label{c3}
 K_0(x) =
      \int_1^{\infty}dt\, e^{-xt}
         \left(\frac1{t^3}+\frac1{2t^5}\right)\,
     \sqrt{t^2-1}\,,
\end{equation}
and the nuclear-charge density $\rho(\bm{r})$ is spherically symmetric and
normalized by the condition $\int d\bm{r}\,\rho(\bm{r}) = 1$. The Wichmann-Kroll
potential is conveniently expressed in terms of the VP charge density
$\rho^{(3+)}_{\rm VP}(r)$,
\begin{equation}\label{c4}
U_{\rm WK}(r) = \int_0^{\infty} dr\pr\, {r\pr}^2\,\frac1{r_>} \,\rho^{(3+)}_{\rm
VP}(r\pr)\,,
\end{equation}
where $r_> = \max (r,r\pr)$. The VP charge density can be
written as
\begin{align}\label{c5}
  \rho^{(3+)}_{\rm VP}(r)  = \frac{2 \alpha}{\pi}\,
      {\rm Re }\
       \sum_{\kappa}|\kappa|
       \int_0^{\infty}d\omega\, {\rm Tr}\  G^{(3+)}_{\kappa}(i\omega,r,r)\,,
\end{align}
where $G^{(3+)}_{\kappa}$ denotes the radial part of the electron propagator that
contains three and more interactions with the binding Coulomb field and $\kappa$
is the relativistic angular quantum number. It was shown
\cite{gyulassy:75,soff:88:vp} that no spurious terms arise in a numerical
evaluation of Eq.~(\ref{c5}) if the expansion over $\kappa$ is terminated by a
finite cutoff parameter.

According to the Furry theorem, all parts of the electron propagator that contain
an even number of interactions with the binding field yield a vanishing
contribution to the VP charge density. Thus, in numerical evaluations, the
function $G^{(3+)}_{\kappa}$ in Eq.~(\ref{c5}) can be substituted by
$G^{(2+)}_{\kappa}$, which is conveniently expressed as
\begin{align}\label{c6}
G^{(2+)}_{\kappa}(\omega,x,y) = \int_0^{\infty}&\, dz\, z^2 \,
G^{(0)}_{\kappa}(\omega,x,z)\,
  V(z)\,
  \nonumber \\ \times &\
  [G_{\kappa}(\omega,z,y)-G^{(0)}_{\kappa}(\omega,z,y)]\,,
\end{align}
where $G_{\kappa}$ is the radial part of the bound-electron propagator,
$G^{(0)}_{\kappa}$ is the radial part of the free electron propagator, and $V(z)$
is the binding potential.

Formulas (\ref{c2})-(\ref{c6}) were employed for the numerical evaluation of the
one-loop VP potential in this work, using the numerical procedure developed in
Refs.~\cite{soff:88:vp,artemyev:97}. The Uehling potential was calculated with the
Fermi model of the nuclear-charge distribution, whereas for the Wichmann-Kroll
potential, the spherical shell model [$\rho(r) \propto \delta(r-R)$] was employed.
The summation over $\kappa$ in Eq.~(\ref{c5}) was extended up to $|\kappa_{\rm
max}| = 10$.

We mention that in the case when high numerical accuracy is not needed and the
finite nuclear size correction may be disregarded, the Wichmann-Kroll potential
can be conveniently evaluated by employing analytical approximation formulas
reported in Ref.~\cite{fainshtein:91}.


\section{Two-loop QED corrections}  \label{sec:twoloop}

The two-loop contributions to the energy shift are conveniently represented in
terms of the dimensionless function $F(\Za)$ defined by
\begin{equation}\label{01}
 \Delta E  = m\,\left(\frac{\alpha}{\pi}\right)^2\,
                 \frac{(Z\,\alpha)^4}{n^3}\,F(Z\,\alpha)\,,
\end{equation}
where $n$ is the principal quantum number.


\subsection{Self-energy in the Coulomb potential modified by the vacuum
polarization}

We start our consideration of the two-loop corrections with the set of the three
diagrams in Fig.~\ref{fig:2order}(d)-(f). This set is gauge invariant and can be
regarded as the one-loop SE correction in the combined field of the nuclear
Coulomb and the VP potential. The corresponding energy shift will be referred to
as the SEVP correction in the following.

Formal expressions for the SEVP correction can be obtained by considering the
first-order perturbation of the one-loop SE correction by the
VP potential (\ref{ii1}). Perturbations of the
reference-state wave function, the binding energy, and the electron propagator
give rise to the irreducible, the reducible, and the vertex contributions,
respectively. The irreducible part is
\begin{equation}
\Delta E_{\rm SEVP}^{\rm ir} = \lbr a| \gamma^0 \Sigma_R(\vare_a)|\delta a\rbr
                 +  \lbr \delta a| \gamma^0 \Sigma_R(\vare_a)|a\rbr\,,
\end{equation}
where $\Sigma_R$ is the renormalized SE operator, and $|\delta a\rbr$
is the first-order perturbation of the reference-state wave function $|a\rbr$
by $U_{\rm VP}$. The reducible part is given by
\begin{equation}
\Delta E_{\rm SEVP}^{\rm red} = \lbr a|U_{\rm VP}|a\rbr\,
        \left.     \lbr a| \gamma_0\,\frac{\partial}{\partial \vare}
         \Sigma_R(\vare)
           |a\rbr  \right|_{\vare=\vare_a} \,,
\end{equation}
and the vertex part is
\begin{align}
\Delta E_{\rm SEVP}^{\rm ver} &\ =
2i\alpha\int_{-\infty}^{\infty}d\omega\, \sum_{n_1n_2} \nonumber \\ &
\times \frac{\lbr n_1|U_{\rm VP}|n_2\rbr\, \lbr
an_2|\alpha_{\mu}\alpha_{\nu}D^{\mu\nu}(\omega)|n_1a\rbr}
{(\vare_a-\omega-\vare_{n_1})(\vare_a-\omega-\vare_{n_2})}\,,
\end{align}
where the summation over $n_{1,2}$ goes over the Dirac spectrum, and the
virtual-state energies are assumed to have a small imaginary addition, $\vare_{n}
\to \vare_{n}(1-i0)$.

The problem of calculating the SE correction in the presence of the perturbing
potential has been extensively studied in the literature, see, e.g.,
Ref.~\cite{indelicato:91:tca,blundell:97:pra,persson:97:g,indelicato:2001:pra}. In
this work, we employ the general scheme developed in our previous study
\cite{yerokhin:05:hfs} for the case of the SE correction to the hyperfine
splitting. Several modifications were introduced into the scheme, among them the
inclusion of the finite nuclear size effect. This modification was essential,
firstly, because the effect is significantly enhanced by the singular behavior of
the VP potential and, secondly, because the extended nuclear charge distribution
removes the logarithmic singularity of the point-nucleus Uehling potential, which
simplifies numerical integrations considerably. Even with the extended nuclear
size, the usage of extremely fine grids was required in the nuclear region, in
order to achieve a high controllable accuracy in radial integrations.

In actual calculations, the Fermi model of the nuclear-charge distribution was
employed for the evaluation of the reference-state wave function and the Uehling
potential, whereas the electron propagator(s) inside the SE loop and the
Wichmann-Kroll potential were calculated with the spherical shell model [$\rho(r)
\propto \delta(r-R)$]. For systems with $Z<10$, we neglected the nuclear-size
dependence in the electron propagators. The values of the root-mean-square (rms)
radii of the nuclear-charge distribution were taken from Ref.~\cite{angeli:04} in
most cases. For uranium, we used the value $\rms = 5.8569\,(33)$~fm obtained in
the recent re-evaluation of experimental data \cite{kozhedub:08}. In the case of
fermium ($Z=100$), there is no experimental results available, so we used the
interpolation formula from Ref.~\cite{johnson:85} and assigned a (conventional)
uncertainty of 1\% to the resulting value. The rms radii used in the present
investigation are listed in second column of Table~\ref{tab:sevp}.

%
%
%
\begin{table*}
\setlength{\LTcapwidth}{\textwidth}
\caption{Energy shifts due the SEVP correction, in units of  $F(\Za)$.
Uncertainties specified include the estimated errors due to the
nuclear charge distribution models and the values of the rms radius. \label{tab:sevp}}
\begin{ruledtabular}
\begin{tabular}{c.....}
     $Z$        & \multicolumn{1}{c}{$\rms$[fm]}
            & \multicolumn{1}{c}{$1s$}
                               & \multicolumn{1}{c}{$2s$}
                                                  & \multicolumn{1}{c}{$2p_{1/2}$}
                                                        & \multicolumn{1}{c}{$2p_{3/2}$}\\
\hline\\[-9pt]
   1  &  0.8x79\,(9)  &  0.01x474\,(2)   &   0.01x459\,(2)  &   -0.000x025      &   -0.000x022      \\
   2  &  1.6x76\,(3)  &  0.02x993\,(2)   &   0.02x945\,(4)  &   -0.000x086      &   -0.000x076      \\
   3  &  2.4x3\,(3)   &  0.04x519\,(3)   &   0.04x430\,(5)  &   -0.000x175      &   -0.000x152      \\
   5  &  2.4x1\,(3)   &  0.07x546\,(4)   &   0.07x355\,(6)  &   -0.000x414      &   -0.000x352      \\
   7  &  2.5x58\,(7)  &  0.10x508\,(3)   &   0.10x208\,(7)  &   -0.000x715      &   -0.000x597      \\
  10  &  3.0x05\,(2)  &  0.14x804\,(3)   &   0.14x351\,(5)  &   -0.001x242      &   -0.001x008      \\
  15  &  3.1x89\,(2)  &  0.21x618\,(3)   &   0.20x961\,(4)  &   -0.002x206      &   -0.001x700      \\
  20  &  3.4x76\,(1)  &  0.28x099\,(5)   &   0.27x394\,(5)  &   -0.003x130      &   -0.002x265      \\
  25  &  3.7x06\,(2)  &  0.34x394\,(4)   &   0.33x841\,(6)  &   -0.003x868      &   -0.002x577      \\
  30  &  3.9x29\,(1)  &  0.40x649\,(5)   &   0.40x497\,(6)  &   -0.004x269\,(1) &   -0.002x529\,(1) \\
  35  &  4.1x63\,(2)  &  0.47x012\,(3)   &   0.47x570\,(6)  &   -0.004x161\,(1) &   -0.002x025\,(1) \\
  40  &  4.2x70\,(1)  &  0.53x642\,(4)   &   0.55x290\,(5)  &   -0.003x332\,(1) &   -0.000x974\,(4) \\
  45  &  4.4x94\,(2)  &  0.60x687\,(3)   &   0.63x898\,(5)  &   -0.001x500\,(1) &    0.000x711\,(2) \\
  50  &  4.6x54\,(1)  &  0.68x338\,(3)   &   0.73x709\,(5)  &    0.001x724\,(2) &    0.003x116\,(6) \\
  55  &  4.8x04\,(5)  &  0.76x802\,(3)   &   0.85x095\,(4)  &    0.006x884\,(2) &    0.006x34\,(1)  \\
  60  &  4.9x12\,(2)  &  0.86x337\,(4)   &   0.98x540\,(4)  &    0.014x770\,(3) &    0.010x449\,(6) \\
  65  &  5.1x\,(2)    &  0.97x23\,(7)    &   1.14x63\,(8)   &    0.026x53\,(2)  &    0.015x548\,(8) \\
  70  &  5.3x11\,(6)  &  1.09x812\,(5)   &   1.34x094\,(6)  &    0.043x840\,(4) &    0.021x71\,(1)  \\
  75  &  5.3x4\,(1)   &  1.24x78\,(2)    &   1.58x27\,(2)   &    0.069x408\,(9) &    0.029x019\,(10)\\
  80  &  5.4x63\,(2)  &  1.42x573\,(5)   &   1.88x34\,(2)   &    0.107x174\,(8) &    0.037x514\,(4) \\
  83  &  5.5x21\,(3)  &  1.55x007\,(6)   &   2.10x10\,(2)   &    0.138x236\,(10)&    0.043x186\,(7) \\
  90  &  5.7x1\,(5)   &  1.90x5\,(2)     &   2.75x1\,(3)    &    0.248x8\,(2)   &    0.057x98\,(3)  \\
  92  &  5.8x57\,(3)  &  2.02x38\,(4)    &   2.97x86\,(8)   &    0.294x00\,(2)  &    0.062x51\,(1)  \\
 100  &  5.8x6\,(6)   &  2.65x0\,(4)     &   4.22x4\,(8)    &    0.587x1\,(8)   &    0.081x83\,(6)  \\
\end{tabular}
\end{ruledtabular}
\end{table*}

The calculational results for the SEVP correction for the $1s$, $2s$, $2p_{1/2}$,
and $2p_{3/2}$ states of H-like ions are presented in Table~\ref{tab:sevp},
expressed in terms of the function $F(\Za)$ defined by Eq.~(\ref{01}). Our results
are in good agreement with the values reported previously for uranium, lead, and
ytterbium in Ref.~\cite{persson:96:pra}.  The uncertainty specified in the table
includes the numerical error and the estimated errors due to the models of the
nuclear charge distribution and due to uncertainties of the rms radii. The model
dependence of the results was estimated by switching between the Fermi, the
uniform, and the spherical-shell models in evaluations of the Uehling potential
and the wave functions; the largest deviation was taken as the error due to the
nuclear model. The error due to the uncertainty of the nuclear radius was obtained
by repeating the calculations with the rms radii varied within the error bars
specified in the table. All three errors were added quadratically. In
Table~\ref{tab:sevp} and in the tables that follow, the omitted uncertainty means
that the expected error is smaller than the last significant digit specified.

\begin{figure}[t]
\centerline{
\resizebox{\columnwidth}{!}{%
  \includegraphics{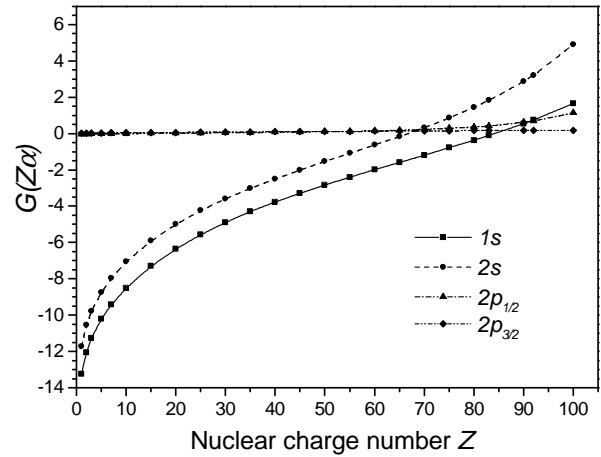}
}}
 \caption{Higher-order remainder $G(\Za)$ for the SEVP correction. Errors due to
the model of the nuclear-charge distribution and the rms radii are not shown.
 \label{fig:sevp}}
\end{figure}

It is instructive to compare our nonpertubative results with the ones obtained
within the $\Za$-expansion approach. The $\Za$ expansion of the SEVP correction
reads
\begin{align}
    \Delta E_{\rm SEVP} = &\,m\,\left(\frac{\alpha}{\pi}\right)^2\, \frac{(\Za)^5}{n^3}\,
     \Bigl\{ a_{50} + (\Za)\,\ln^2\left[(\Za)^{-2}\right]\,a_{62}
 \nonumber \\ &
     + (\Za)\,\ln\left[(\Za)^{-2}\right]\,a_{61}+
     (\Za)\,G(\Za)\Bigr\}\,,
\end{align}
where the function $G(\Za)$ is the higher-order remainder, $G(0\alpha)=a_{60}$\,.
The results known for the coefficients of this expansion are:
\cite{pachucki:93:pra,eides:93,karshenboim:96:jetp,pachucki:01:pra,jentschura:05:sese}:
\begin{align}
  a_{50} =&\ 1.920576\,\delta_{l,0}\,,\\
  a_{62} =&\ \frac{4}{45}\,\delta_{l,0}\,,\\
  a_{61} =&\ \frac{16}{15}\,\left(\frac29+\ln 2\right)\,\delta_{l,0}
    \nonumber \\
         &\ -\frac{32}{45}\left( \frac34 +\frac1{4n^2}-\frac1n +\gamma
               +\Psi(n)-\ln n \right)\,\delta_{l,0}
    \nonumber \\
         &\ -\frac{8}{135}\frac{n^2-1}{n^2}\,\delta_{l,1}\,.
\end{align}
The higher-order remainder $G(\Za)$  was inferred from our numerical results for
the SEVP correction and plotted on Fig.~\ref{fig:sevp}. For hydrogen, the results
of our direct numerical evaluation are: $G_{1s}(\alpha) = -13.2\,(4)$,
$G_{2s}(\alpha) = -11.7\,(4)$, $G_{2p_{1/2}}(\alpha) = -0.034$, and
$G_{2p_{3/2}}(\alpha) = 0.015$. For the normalized difference of the $2s$ and
$1s$ states and for the fine-structure difference, these values are consistent
with the analytical results $a_{60}(2s)-a_{60}(1s) = 1.491199$ and
$a_{60}(2p_{3/2})-a_{60}(2p_{1/2}) = 1/20$ \cite{jentschura:05:sese}.

\subsection{Two-loop vacuum polarization}

The two-loop VP correction is represented by the diagrams in
Fig.~\ref{fig:2order}(g)-(i). (The term the ``two-loop VP'' will be abbreviated as
``VPVP'' in the following.) It will be convenient to split our evaluation of the
VPVP correction into two parts, considering separately the diagram (g) and the
remaining two diagrams (h) and (i).

\subsubsection{Diagram (g)}

The correction induced by the diagram in Fig.~\ref{fig:2order}(g) can be regarded
as the second-order perturbation contribution induced by the one-loop VP potential
$U_{\rm VP}$,
\begin{equation}\label{c1}
 \Delta E_{{\rm VPVP},\,g} = \sum_{n\neq a} \frac{\lbr a| U_{\rm VP}|n\rbr\,
   \lbr n| U_{\rm VP}|a\rbr}{\vare_n-\vare_a}\,.
\end{equation}
The numerical evaluation of this correction is relatively simple. It was carried
out by employing  the general scheme developed for the VP potential and described
in Sec.~\ref{sec:vp}. The summation over the Dirac spectrum was performed by the
dual-kinetic-balance basis set method \cite{shabaev:04:DKB}.

The numerical values of the energy shifts induced by the diagram in
Fig.~\ref{fig:2order}(g) are presented in Table~\ref{tab:vpvp} for the $n=1$ and
$n=2$ states of H-like ions. The results are expressed in terms of the function
$F(\Za)$ defined by Eq.~(\ref{01}). Good agreement is found with the previous
evaluations of this corrections \cite{persson:96:pra,beier:97:jpb}. Our
calculation accounts for the extended nuclear charge distribution (with the Fermi
model employed for the Uehling potential and the wave functions and the spherical
shell model, for the Wichmann-Kroll potential). The uncertainties listed in the
table include the numerical error as well as the estimated errors due to the
models of the nuclear charge distribution and due to the uncertainties of the
values of the rms radii. The estimation of errors was done similarly to that for
the SEVP correction.

\begin{figure}[t]
\centerline{
\resizebox{\columnwidth}{!}{%
  \includegraphics{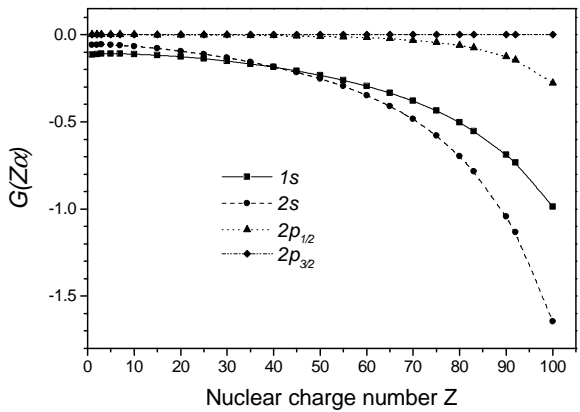}
}}
 \caption{Higher-order remainder $G(\Za)$ induced
by the diagram in Fig.~\ref{fig:2order}(g).
 \label{fig:vpvp}}
\end{figure}

The higher-order part of the correction can be identified by taking into account
its $\Za$ expansion of the form
\begin{align}\label{c7}
 \Delta E_{{\rm VPVP},\,g} &\ = m\,\left( \frac{\alpha}{\pi}\right)^2
 \frac{(\Za)^5}{n^3}
     \biggl\{ a_{50}
\nonumber \\ &
     + (\Za)\,\ln\left[(\Za)^{-2}\right]\,a_{61}
         + (\Za)\, G(\Za) \biggr\}\,,
\end{align}
where \cite{pachucki:93:pra,eides:92,pachucki:01:pra}
\begin{eqnarray}\label{c7a}
a_{50} &=& -\frac{23\pi}{1134}\,\delta_{l,0}\,,\\
a_{61} &=& -\frac{8}{225}\,\delta_{l,0}\,,
\end{eqnarray}
and $G(\Za)$ is the higher-order remainder, $G(0\alpha)=a_{60}$. The remainder
$G(\Za)$ inferred from our numerical results is plotted on Fig.~\ref{fig:vpvp}.
The results for hydrogen are: $G_{1s}(\alpha) = -0.115$, $G_{2s}(\alpha) =
-0.059$, $|G_{2p_{1/2}}(\alpha)| < 10^{-4}$, and $|G_{2p_{3/2}}(\alpha)| <
10^{-4}$.

%
%
\begin{table*}
\setlength{\LTcapwidth}{\textwidth}
 \caption{Energy shifts induced by the VPVP
diagram in Fig.~\ref{fig:2order}(g), in units of $F(\Za)$. The uncertainties
specified include the estimated errors due to the nuclear charge distribution
models and due to the rms radii. \label{tab:vpvp}}
\begin{ruledtabular}
\begin{tabular}{c....}
     $Z$        & \multicolumn{1}{c}{$1s$}
                               & \multicolumn{1}{c}{$2s$}
                                                  & \multicolumn{1}{c}{$2p_{1/2}$}
                                                        & \multicolumn{1}{c}{$2p_{3/2}$}\\
\hline\\[-9pt]
   1   &   -0.000x490       &  -0.000x487      &                   &              \\
   2   &   -0.001x018       &  -0.001x006      &                   &              \\
   3   &   -0.001x578       &  -0.001x552      &                   &              \\
   5   &   -0.002x784       &  -0.002x716      &                   &              \\
   7   &   -0.004x092       &  -0.003x967      &   -0.00x0001      &              \\
  10   &   -0.006x231       &  -0.005x995      &   -0.00x0002      &              \\
  15   &   -0.010x260\,(1)  &  -0.009x803\,(1) &   -0.00x0009      &              \\
  20   &   -0.014x896\,(1)  &  -0.014x216\,(1) &   -0.00x0026      &   -0.00x0001 \\
  25   &   -0.020x225\,(2)  &  -0.019x375\,(2) &   -0.00x0060      &   -0.00x0002 \\
  30   &   -0.026x376\,(3)  &  -0.025x480\,(3) &   -0.00x0124      &   -0.00x0003 \\
  35   &   -0.033x524\,(5)  &  -0.032x796\,(5) &   -0.00x0235      &   -0.00x0005 \\
  40   &   -0.041x928\,(8)  &  -0.041x708\,(8) &   -0.00x0421      &   -0.00x0007 \\
  45   &   -0.051x85\,(1)   &  -0.052x65\,(1)  &   -0.00x0727      &   -0.00x0010 \\
  50   &   -0.063x74\,(2)   &  -0.066x31\,(2)  &   -0.00x1218\,(1) &   -0.00x0015 \\
  55   &   -0.078x10\,(3)   &  -0.083x56\,(3)  &   -0.00x1998\,(1) &   -0.00x0021 \\
  60   &   -0.095x68\,(5)   &  -0.105x66\,(6)  &   -0.00x3233\,(2) &   -0.00x0029 \\
  65   &   -0.117x3\,(2)    &  -0.134x2\,(3)   &   -0.00x518\,(1)  &   -0.00x0039 \\
  70   &   -0.144x1\,(1)    &  -0.171x3\,(1)   &   -0.00x8238\,(7) &   -0.00x0051 \\
  75   &   -0.178x5\,(2)    &  -0.221x2\,(2)   &   -0.01x313\,(1)  &   -0.00x0067 \\
  80   &   -0.222x1\,(3)    &  -0.288x0\,(4)   &   -0.02x094\,(3)  &   -0.00x0085 \\
  83   &   -0.254x2\,(4)    &  -0.339x3\,(5)   &   -0.02x778\,(4)  &   -0.00x0099 \\
  90   &   -0.352x2\,(9)    &  -0.505x\,(1)    &   -0.05x43\,(1)   &   -0.00x0136 \\
  92   &   -0.386x5\,(9)    &  -0.566x\,(1)    &   -0.06x58\,(1)   &   -0.00x0149 \\
 100   &   -0.583x\,(3)     &  -0.935x\,(4)    &   -0.14x82\,(6)   &   -0.00x0209 \\
\end{tabular}
\end{ruledtabular}
\end{table*}

\subsubsection{Diagrams (h) and (i)}

We now consider the energy shift induced by the diagrams in
Fig.~\ref{fig:2order}(h) and (i). Its leading (in $\Za$) part can be obtained
within the {\em free-loop} approximation, i.e., keeping the first nonvanishing
term in the expansion of the fermion loop in terms of the binding potential. The
corresponding expression was derived long ago in the classical paper by
K\"all\'{e}n and Sabry \cite{kaellen:55}. Later, it was re-derived by a number of
other techniques \cite{schwinger:particles,barbieri:73,broadhurst:93}. According
to the Furry theorem, corrections to the K\"all\'{e}n-Sabry potential are
suppressed by a factor of $(\Za)^2$, so that this potential is supposed to yield a
dominant contribution even in the medium-$Z$ region. In the present investigation,
diagrams (h) and (i) will be approached within the free-loop approximation only.
In order to stress this fact, we will use the superscript ``KS'' in the formulas
below.

In the free-loop approximation, the correction induced by the diagrams in
Fig.~\ref{fig:2order}(h) and (i) is given by the expectation value of the
K\"all\'{e}n-Sabry potential,
\begin{equation}\label{a0}
    \Delta E_{{\rm VPVP},\,hi}^{KS} = \lbr a | V_{KS}|a\rbr\,.
  \end{equation}
For a spherically symmetric nuclear charge distribution, the K\"all\'{e}n-Sabry
potential can be conveniently written in the form \cite{fullerton:76}
\begin{widetext}
\begin{equation}\label{a1}
    V_{KS}(r) = -\frac{\alpha^3 Z}{m \pi r}\,
    \int_0^{\infty}dr\pr\,r\pr \rho(r\pr)\,
      \left[ L_0(2m|r-r\pr|)- L_0(2m|r+r\pr|)\right]\,,
\end{equation}
where
\begin{eqnarray}\label{a2}
  L_0(x) &=& \int_1^{\infty}dt\, \frac{e^{-xt}}{t}
    \left\{ \left(-\frac{13}{54t^2}-\frac{7}{108t^4}-\frac{2}{9t^6}\right)\,
      \sqrt{t^2-1} +
      \left(\frac{44}{9t}-\frac{2}{3t^3}-\frac{5}{4t^5}-\frac{2}{9t^7}\right)\,
      \ln [t+\sqrt{t^2-1}]
        \right.
 \nonumber \\ &&
  + \left(-\frac{4}{3t^2}-\frac{2}{3t^4}\right)\,\sqrt{t^2-1}\,\ln [8t(t^2-1)]
 \nonumber \\ &&
  \left.
 + \left( \frac{8}{3t}-\frac{2}{3t^5}\right)
   \int_t^{\infty}dy\,\left(\frac{3y^2-1}{y(y^2-1)}\,\ln[y+\sqrt{y^2-1}]-
      \frac1{\sqrt{y^2-1}}\ln[8y(y^2-1)] \right)
      \right\}\,,
\end{eqnarray}
\end{widetext}
and the nuclear charge density is normalized by $\int d\bm{r}\,\rho(\bm{r}) =
1\,$.

The results of our numerical evaluation of the energy shift due to the
K\"all\'{e}n-Sabry potential are listed in Table~\ref{tab:KS}. The numerical
values presented agree well with the results of the previous studies
\cite{beier:88,schneider:93:ks}. Our calculation was carried out with employing
the Fermi model for the nuclear charge distribution. The values for the rms radii
and their uncertainties are listed in Table~\ref{tab:sevp}. The uncertainties
specified in Table~\ref{tab:KS} for our numerical results include the numerical
error as well as the errors due to the nuclear-charge distribution model and due
to uncertainties of the rms radii. The error due to the nuclear model was
estimated by taking the difference of the results obtained with the Fermi and the
uniform model.

\begin{figure}[t]
\centerline{
\resizebox{\columnwidth}{!}{%
  \includegraphics{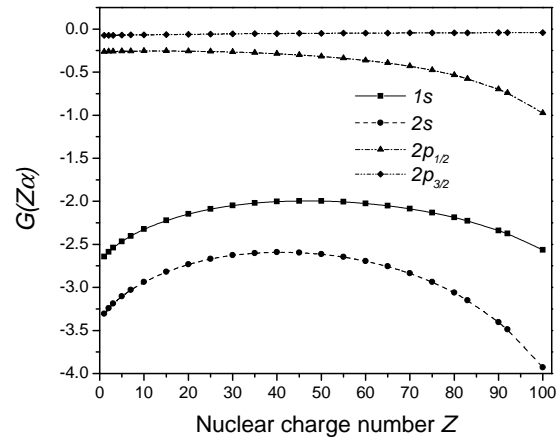}
}}
 \caption{Higher-order remainder $G(\Za)$ induced by
the K\"all\'{e}n-Sabry contribution.
 \label{fig:KS}}
\end{figure}

The $\Za$ expansion of the K\"all\'{e}n-Sabry  contribution reads
\begin{align} \label{a3}
     \Delta E_{{\rm VPVP},\, hi}^{KS}
         = &\,m\left(\frac{\alpha}{\pi}\right)^2\, \frac{(\Za)^4}{n^3}\,
     \Bigl\{ a_{40}+ (\Za)\,a_{50}
 \nonumber \\ &
     + (\Za)^2\,\ln\left[(\Za)^{-2}\right]\,a_{61}+
     (\Za)^2\,G(\Za)\Bigr\}\,,
\end{align}
where the function $G(\Za)$ is the higher-order remainder. The results for the
first terms of the $\Za$ expansion are
\cite{baranger:52,appelquist:70,pachucki:93:pra,eides:92,pachucki:01:pra}:
\begin{eqnarray}
  a_{40} &=& -\frac{82}{81}\,\delta_{l0}\,,\\
  a_{50} &=& \left(
   \frac{45331}{39690} - \frac{25}{63}\pi +\frac{52}{63} \ln 2\right)
   \,\pi\,\delta_{l0}\,,\\
  a_{61} &=& \frac{a_{40}}{2}\,.
\end{eqnarray}
The higher-order remainder $G(\Za)$ inferred from the K\"all\'{e}n-Sabry
contribution is plotted as a function of the nuclear charge number $Z$ on
Fig.~\ref{fig:KS}. For hydrogen, the results for the K\"all\'{e}n-Sabry remainder
term are: $G_{1s}(\alpha) = -2.642$, $G_{2s}(\alpha) = -3.303$,
$G_{2p_{1/2}}(\alpha) = -0.263$, and $G_{2p_{3/2}}(\alpha) = -0.073$.

Our numerical results for the higher-order remainder for the {\it total} VPVP
correction exhibit good agreement with the analytical values obtained in
Ref.~\cite{jentschura:05:sese} for the normalized difference of the $2s$ and $1s$
states and for the fine-structure difference. Indeed, for the VPVP remainder, our
calculation yields: $G_{2s}(\alpha)-G_{1s}(\alpha) = -0.605$ and
$G_{2p_{3/2}}(\alpha)-G_{2p_{1/2}}(\alpha) = 0.190$, to be compared with the
analytical results of $a_{60}(2s)-a_{60}(1s)=-0.611365$ and
$a_{60}(2p_{3/2})-a_{60}(2p_{1/2})= 0.189815$, correspondingly.

A complete evaluation of the VPVP correction beyond the free-loop approximation is
a difficult problem [especially, for diagram (i)] and has not been carried out up
to now. For diagram (h), the calculation is easier and can be performed by a
generalization of methods developed for the one-loop VP correction. For uranium
and lead, such a calculation was reported in Ref.~\cite{plunien:98:epj}. The
numerical values obtained for this diagram for the contribution beyond the
free-loop approximation turned out to be rather small; it is expected that the
corresponding contribution from diagram (i) is much larger. In the absence of a
direct calculation, we estimate the theoretical uncertainty of the VPVP correction
due to the omitted part beyond the free-loop approximation by multiplying the
absolute value of the K\"all\'{e}n-Sabry contribution by a factor of $(\Za)^2$.

%
%
\begin{table*}
\setlength{\LTcapwidth}{\textwidth}
 \caption{Energy shifts due to the K\"all\'{e}n-Sabry potential, in units of $F(\Za)$.
The uncertainties specified include the estimated errors due to the nuclear charge
distribution models and due to the rms radii.  \label{tab:KS} }
\begin{ruledtabular}
\begin{tabular}{c....}
     $Z$        & \multicolumn{1}{c}{$1s$}
                               & \multicolumn{1}{c}{$2s$}
                                                  & \multicolumn{1}{c}{$2p_{1/2}$}
                                                        & \multicolumn{1}{c}{$2p_{3/2}$}\\
\hline\\[-9pt]
   1   &   -1.002x032      &   -1.002x067      &   -0.000x014       &  -0.00x0004     \\
   2   &   -0.992x369      &   -0.992x509      &   -0.000x056       &  -0.00x0015     \\
   3   &   -0.983x258      &   -0.983x568      &   -0.000x125       &  -0.00x0034     \\
   5   &   -0.966x493      &   -0.967x341      &   -0.000x344       &  -0.00x0093     \\
   7   &   -0.951x435      &   -0.953x071      &   -0.000x670       &  -0.00x0179     \\
  10   &   -0.931x629      &   -0.934x898      &   -0.001x360       &  -0.00x0354     \\
  15   &   -0.904x993\,(1) &   -0.912x148      &   -0.003x055       &  -0.00x0757     \\
  20   &   -0.885x141\,(1) &   -0.897x624\,(1) &   -0.005x471       &  -0.00x1287     \\
  25   &   -0.871x198\,(1) &   -0.890x494\,(1) &   -0.008x683       &  -0.00x1927     \\
  30   &   -0.862x607\,(2) &   -0.890x315\,(2) &   -0.012x801       &  -0.00x2669     \\
  35   &   -0.859x040\,(3) &   -0.896x935\,(3) &   -0.017x978       &  -0.00x3504     \\
  40   &   -0.860x403\,(4) &   -0.910x519\,(5) &   -0.024x416       &  -0.00x4426     \\
  45   &   -0.866x633\,(6) &   -0.931x345\,(7) &   -0.032x384       &  -0.00x5431     \\
  50   &   -0.877x972\,(8) &   -0.960x119\,(9) &   -0.042x236       &  -0.00x6517     \\
  55   &   -0.894x73\,(1)  &   -0.997x76\,(1)  &   -0.054x441       &  -0.00x7684     \\
  60   &   -0.917x47\,(2)  &   -1.045x61\,(2)  &   -0.069x624\,(1)  &  -0.00x8931     \\
  65   &   -0.946x8\,(3)   &   -1.105x3\,(4)   &   -0.088x62\,(1)   &  -0.01x0259     \\
  70   &   -0.983x46\,(4)  &   -1.178x95\,(4)  &   -0.112x569\,(2)  &  -0.01x1673     \\
  75   &   -1.029x57\,(7)  &   -1.270x69\,(9)  &   -0.143x102\,(5)  &  -0.01x3175     \\
  80   &   -1.086x01\,(7)  &   -1.383x69\,(9)  &   -0.182x409\,(7)  &  -0.01x4770     \\
  83   &   -1.125x95\,(8)  &   -1.464x2\,(1)   &   -0.211x50\,(1)   &  -0.01x5775     \\
  90   &   -1.240x2\,(5)   &   -1.698x8\,(8)   &   -0.301x71\,(8)   &  -0.01x8267     \\
  92   &   -1.278x4\,(2)   &   -1.779x1\,(2)   &   -0.334x88\,(3)   &  -0.01x9020     \\
 100   &   -1.476x\,(1)    &   -2.201x\,(2)    &   -0.520x2\,(3)    &  -0.02x2227     \\
\end{tabular}
\end{ruledtabular}
\end{table*}

\subsection{Vacuum-polarization insertion into the self-energy photon line}

In this section, we address the correction induced by the SE diagram with the VP
insertion into the photon line, depicted by Fig.~\ref{fig:2order}(k). The
corresponding shift of the energy will be referred to as the S(VP)E correction.
The leading (in $\Za$) part of this correction can be again obtained within the
free-loop approximation. Unlike the VPVP contribution, however, corrections to the
free-loop approximation in this case are suppressed only by the first power of
$\Za$. The leading term beyond this approximation is known from perturbative
calculations \cite{pachucki:93:pra,eides:95:pra}. An all-order calculation of the
S(VP)E correction beyond the free-loop approximation is a difficult problem and
has not been performed so far. In the present investigation, we will approach the
S(VP)E correction within the free-loop approximation only.

The evaluation of the S(VP)E correction within the free-loop approximation can be
performed by a generalization of the method for the one-loop SE correction
described in Section~\ref{sec:se}. The S(VP)E correction is represented by a sum
of the zero-, one-, and many-potential terms:
\begin{equation}\label{b00}
 \Delta E_{\rm SVPE,a} = \Delta E_{\rm SVPE,a}^{(0)}+
             \Delta E_{\rm SVPE,a}^{(1)}+
                        \Delta E_{\rm SVPE,a}^{(2+)}\,,
\end{equation}
where the subscript ``$a$'' indicates the free-loop approximation. Formulas for
the three terms above can be obtained from the corresponding expressions in
Section~\ref{sec:se} by replacing the standard photon propagator by the
``dressed'' one, obtained by inserting the renormalized one-loop VP tensor into
the photon line. In $D = 4-2\eps$ dimensions, the replacement is given by
\cite{adkins:98:cjp,mallampalli:96}
\begin{align} \label{b01}
    \frac1{k^2+i0} \to -\frac{\alpha C_{\eps}}{4\pi}\,m^{2\eps}
      \int_0^1dz\,
        \frac{z^2(1-z^2/3)}{[m^2-k^2(1-z^2)/4-i0]^{1+\eps}}\,,
\end{align}
where $C_{\eps} = (4\pi)^{\eps}\,\Gamma(1+\eps)\,\mu^{2\eps}/m^{2\eps}$.

The zero-potential term is represented by
\begin{equation}\label{b02}
    \Delta E_{\rm SVPE,a}^{(0)}= \int \frac{d\bfp}{(2\pi)^3} \,\,
        \overline{\psi}_a(\bfp)\,  \Sigma_{\rm VP,R}^{(0)}
         (\vare_a,\bfp)\, \psi_a(\bfp)\, ,
\end{equation}
where the free dressed SE operator $\Sigma_{\rm VP}^{(0)}$ is obtained from
Eq.~(\ref{i11}) by the substitution (\ref{b01}), and its renormalized part
$\Sigma^{(0)}_{\rm VP, R}$ is the finite part of the difference $\Sigma^{(0)}_{\rm
VP}-\delta m$ when $\eps$ approaches zero. Applying the standard technique for the
evaluation of momentum integrals (see, e.g., Appendices of
Ref.~\cite{yerokhin:03:epjd}), the difference $\Sigma^{(0)}_{\rm VP}-\delta m$ is
conveniently represented as
\begin{widetext}
\begin{eqnarray}\label{b04}
  \Sigma^{(0)}_{\rm VP}(p)-\delta m &=& \left( \frac{\alpha C_{\eps}}{4\pi}
  \right)^2
    \left[ -\frac{2}{3\eps^2}+\frac{13}{9\eps}-\frac{599}{54}+\frac{8\pi^2}{9}
    \right](\cross{p}-m)
      \nonumber \\ &&
  + \left( \frac{\alpha}{4\pi}\right)^2 8 \int_0^1 dx\, dz\,
   \frac{z^2(1-z^2/3)}{1-z^2} \bigl[ (1-x)\cross{p}-2m \bigr]\,
     \ln \frac{x(1-z^2)Y+4(1-x)}{x^2(1-z^2)+4(1-x)}\,,
\end{eqnarray}
\end{widetext}
where $Y = x(1-\rho)+\rho$ and $\rho = (m^2-p^2)/m^2$.

The one-potential term is given by
\begin{eqnarray}\label{b05}
    \Delta E_{\rm SVPE,a}^{(1)} &=&
        \int \frac{d\bfp_1}{(2\pi)^3}\,
         \frac{d\bfp_2}{(2\pi)^3} \,\,
        \overline{\psi}_a(\bfp_1)\,
 \nonumber \\ && \times
        \Gamma^0_{\rm VP,R}(\vare_a,\bfp_1;\vare_a,\bfp_2)\,
        V_C(\bfq)\, \psi_a(\bfp_2) \, ,
\end{eqnarray}
where $\Gamma^0_{\rm VP,R}$ is the finite (when $\eps\to 0$) part of the time
component of the dressed vertex operator $\Gamma^{\alpha}_{\rm VP}$, which is
obtained from the one-loop vertex operator $\Gamma^{\alpha}$, Eq.~(\ref{i12}), by
the substitution (\ref{b01}). Evaluating the momentum integrations, we obtain the
following representation for the operator $\Gamma^{\alpha}_{\rm VP}$,
\begin{widetext}
\begin{align}\label{b07}
 \Gamma^{\alpha}_{\rm VP}(p_1,p_2) =& \left(\frac{\alpha C_{\eps}}{4\pi} \right)^2
    \, \gamma^{\alpha}\, \left( \frac{2}{3\eps^2}-\frac{13}{9\eps}
          -\frac{877}{54} +\frac{16\pi^2}{9}\right)
  \nonumber \\
 -& \left(\frac{\alpha }{4\pi} \right)^2 \, \int_0^1 dx\,dy\,dz\,
   xz^2\left(1-\frac{z^2}{3}\right) \left\{
      \frac{4\,N^{\alpha}(xb)}{x(1-z^2) D+4m^2(1-x)}
       + \frac{8\,\gamma^{\alpha}}{1-z^2}\,
         \ln \frac{x(1-z^2)D/m^2+4(1-x)}{x^2(1-z^2)+4(1-x)}
          \right\}\,,
\end{align}
\end{widetext}
where $D = xb^2+ m^2 -yp_1^2 -(1-y)p_2^2$, $b = yp_1+(1-y)p_2$, and $
 N^{\alpha}(k) = \gamma_{\sigma}(\cross{p}_1-\cross{k}+m) \gamma^{\alpha}
        (\cross{p}_2-\cross{k}+m) \gamma^{\sigma}\,.$ Using the Ward identity, it is
easy to check that the divergent parts of Eqs.~(\ref{b07}) and (\ref{b04}) (i.e.,
terms $\sim \eps^{-1}$ and $\eps^{-2}$) cancel each other in the total expression
for the energy shift. This is the justification of our definition of the
renormalized parts of the operators $\Sigma^{(0)}_{\rm VP}$ and
$\Gamma^{\alpha}_{\rm VP}$, which consists just in dropping out the divergent part
of Eqs.~(\ref{b04}) and (\ref{b07}).

The numerical evaluation of the zero- and one-potential terms is similar to that
for the one-loop SE correction but is more time consuming, due to the presence of
additional integrations over Feynman parameters. In order to achieve high
numerical accuracy for the one-potential term in the low-$Z$ region, we had to
identify the contribution of Eq.~(\ref{b07}) at $p_1=p_2$ and evaluate it
separately. The subtraction of this contribution in Eq.~(\ref{b05}) makes the
integrand to be a smooth function at $q=0$, which simplifies numerical
integrations considerably.

The many-potential term is given by
\begin{align} \label{b08}
\Delta E_{\rm SVPE,a}^{(2+)}  =& \ 2i \alpha \int_{-\infty}^{\infty} d\omega
        \int d\bfx_1\,
        d \bfx_2\, D^{\mu \nu}_{VP}(\omega, \bfx_{12})\,
  \nonumber \\  & \times
        \psi^{\dag}_a(\bfx_1)\, \alpha_{\mu}\,
        G^{(2+)}(\vare_a-\omega,\bfx_1,\bfx_2)\, \alpha_{\nu}\,
        \psi_a(\bfx_2)\, .
        \nonumber \\
\end{align}
This formula differs from the expression for the many-potential part of the
one-loop SE correction only by the dressed photon propagator $D^{\mu\nu}_{VP}$,
which reads
\begin{align}\label{b09}
    D^{\mu\nu}_{VP}(\omega,\bfx_{12}) = \frac{\alpha}{\pi}\,
      \int_1^{\infty}dt\, &\ \sqrt{t^2-1}\,\, \frac{2t^2+1}{3t^4}\,
       \nonumber \\ &\times
       D^{\mu\nu}(\omega,\bfx_{12};2mt)\,.
\end{align}
$D^{\mu\nu}(\omega,\bfx;\lambda)$ is the propagator of the photon with a mass
$\lambda$, whose expression in the Feynman gauge is
\begin{equation}\label{b010}
 D^{\mu\nu}(\omega,{\bfx}_{12};\lambda) = g^{\mu\nu}\,
   \frac{\exp[i\sqrt{\omega^2-\lambda^2+i0}\,x_{12}]}{4\pi\,x_{12}}\,.
\end{equation}

It is easy to see that the numerical calculation of the many-potential term falls
naturally into two steps: (i) the evaluation of the many-potential part of the
one-loop SE correction with an effective photon mass $\lambda=2 m t$ and (ii) the
numerical integration over $t$ as given by Eq.~(\ref{b09}). There is even a
certain simplification as compared to the one-loop case. It is a common approach
to deform the contour of the $\omega$ integration in Eq.~(\ref{b08}), separating
it into the low-energy and the high-energy part (see, e.g.,
Ref.~\cite{yerokhin:99:pra} for details). In the case of the S(VP)E correction,
the contribution induced by the low-energy part of the contour vanishes
identically, which is due to the condition on the effective photon mass $\lambda
\geq 2m$.

The results of our numerical evaluation of the S(VP)E correction for the $n=1$ and
$n=2$ states of H-like ions are presented in Table~\ref{tab:SVPE}. Our numerical
results are in good agreement with the data obtained previously for the $1s$ state
\cite{mallampalli:96,persson:96:pra} and with the $2s$ and $2p_{1/2}$ values for
$Z=92$ from Ref.~\cite{persson:96:pra}. Our calculation was performed within the
free-loop approximation and for the point nuclear model. The uncertainty specified
in the table is the numerical error only. We estimate the theoretical uncertainty
due to uncalculated terms beyond the free-loop approximation by multiplying the
absolute value of the correction by a factor of $3\,(\Za)$. This factor arises as
a ratio of the leading-order contribution beyond the free-loop approximation for
the $1s$ state, $-0.386\, m(\alpha/\pi)^2(\Za)^5$ \cite{pachucki:93:pra}, and the
leading-order contribution within this approximation, $0.142\,
m(\alpha/\pi)^2(\Za)^4$ \cite{appelquist:70}.

The inclusion of the finite nuclear size (FNS) effect is not necessary at present
for the S(VP)E correction, since it is expected to yield a much smaller
contribution than the error due to the free-loop approximation. The relative
contribution of the FNS effect on the S(VP)E correction can be estimated by taking
the relative values of this effect for the one-loop SE correction. To the leading
orders in $\Za$ and $\ln R$, the relative value of the FNS-SE effect for the $ns$
states is \cite{milstein:02:prl,milstein:04}
\begin{align}
 \delta_{FNS} &\ = -\alpha\left[\Za\left(\frac{23}{4}-4\ln 2\right)
   \right.
    \nonumber \\ &
    \left.
    + \frac{(\Za)^2}{\pi}\left( \frac{15}{4}-\frac{\pi^2}{6}\right)
        \ln (b/R) \right]\,,
\end{align}
where $b = \exp[1/(2\gamma)-C-5/6]\,$, $\gamma = \sqrt{1-(\Za)^2}$, $C\approx
0.557$ is the Euler constant, and $R$ is the nuclear radius. Analogous formulas
for the $np_{1/2}$ and $np_{3/2}$ states can be found in Ref.~\cite{milstein:04}.
Numerical values of $\delta_{FNS}$ are within 3\% for the whole region of the
nuclear charge numbers.

%
%
\begin{table}
\caption{Energy shifts due the S(VP)E correction evaluated within the free-loop
approximation and for the point nucleus, in units of  $F(\Za)$. \label{tab:SVPE} }
\begin{ruledtabular}
\begin{tabular}{c....}
     $Z$        & \multicolumn{1}{c}{$1s$}
                               & \multicolumn{1}{c}{$2s$}
                                                  & \multicolumn{1}{c}{$2p_{1/2}$}
                                                        & \multicolumn{1}{c}{$2p_{3/2}$}\\
\hline\\[-9pt]
  1 &     0.1404x59\,(4)   &     0.1404x65\,(3)   &    -0.0052x28\,(2)   &     0.0026x15\,(2) \\
  2 &     0.1390x02\,(4)   &     0.1390x26\,(2)   &    -0.0052x26\,(1)   &     0.0026x15\,(1) \\
  3 &     0.1376x50\,(4)   &     0.1377x05\,(2)   &    -0.0052x23\,(1)   &     0.0026x15\,(1) \\
  5 &     0.1352x19\,(4)   &     0.1353x70\,(2)   &    -0.0052x12\,(1)   &     0.0026x14\,(1) \\
  7 &     0.1330x99\,(3)   &     0.1333x90\,(4)   &    -0.0051x96\,(3)   &     0.0026x12\,(3) \\
 10 &     0.1304x2\,(1)    &     0.1310x0\,(1)    &    -0.0051x6         &     0.0026x0      \\
 15 &     0.1270x8\,(1)    &     0.1283x5\,(1)    &    -0.0050x7         &     0.0025x9       \\
 20 &     0.1249x3\,(1)    &     0.1271x5\,(1)    &    -0.0049x2         &     0.0025x8       \\
 25 &     0.1238x1         &     0.1272x7         &    -0.0047x1         &     0.0025x5       \\
 30 &     0.1236x6         &     0.1286x6         &    -0.0044x2         &     0.0025x2       \\
 35 &     0.1244x4         &     0.1313x4         &    -0.0040x3         &     0.0024x9       \\
 40 &     0.1261x5         &     0.1353x6\,(1)    &    -0.0035x0         &     0.0024x6       \\
 45 &     0.1288x2         &     0.1408x5\,(1)    &    -0.0028x0         &     0.0024x2       \\
 50 &     0.1325x4         &     0.1480x1\,(1)    &    -0.0018x7         &     0.0023x7       \\
 55 &     0.1374x2         &     0.1571x1\,(1)    &    -0.0006x4         &     0.0023x2       \\
 60 &     0.1436x2         &     0.1685x1\,(1)    &     0.0009x8         &     0.0022x6       \\
 65 &     0.1513x7         &     0.1827x1\,(1)    &     0.0031x4         &     0.0022x0       \\
 70 &     0.1609x9\,(1)    &     0.2004x2\,(1)    &     0.0060x3         &     0.0021x3       \\
 75 &     0.1729x1\,(1)    &     0.2226x0\,(1)    &     0.0099x4         &     0.0020x5       \\
 80 &     0.1877x3\,(1)    &     0.2505x9\,(2)    &     0.0152x8         &     0.0019x7       \\
 83 &     0.1983x5\,(1)    &     0.2709x7\,(2)    &     0.0194x3         &     0.0019x1       \\
 90 &     0.2298x7\,(2)    &     0.3328x5\,(3)    &     0.0332x2         &     0.0017x7       \\
 92 &     0.2411x0\,(2)    &     0.3553x5\,(3)    &     0.0386x3         &     0.0017x2       \\
100 &     0.3006x4\,(3)    &     0.4784x2\,(6)    &     0.0712x4\,(1)    &
0.0015x2
\end{tabular}
\end{ruledtabular}
\end{table}

\begin{figure}[t]
\centerline{
\resizebox{\columnwidth}{!}{%
  \includegraphics{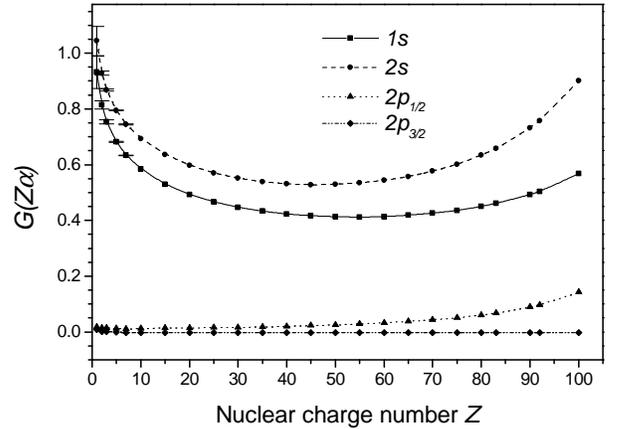}
}}
 \caption{Higher-order remainder $G(\Za)$ for the S(VP)E correction within the free-loop
approximation.
 \label{fig:svpe}}
\end{figure}

The $\Za$ expansion of the S(VP)E correction within the free-loop approximation
reads
\begin{align} \label{b10}
    \Delta E_{\rm SVPE,a} = &\,m\,\left(\frac{\alpha}{\pi}\right)^2\, \frac{(\Za)^4}{n^3}\,
     \Bigl\{ a_{40}+ (\Za)\,a_{50}
 \nonumber \\ &
     + (\Za)^2\,\ln\left[(\Za)^{-2}\right]\,a_{61}+
     (\Za)^2\,G(\Za)\Bigr\}\,,
\end{align}
where the function $G(\Za)$ is the higher-order remainder. The results for the
first terms of the $\Za$ expansion are
\cite{baranger:52,appelquist:70,pachucki:93:pra,eides:95:pra,pachucki:01:pra}:
\begin{align}
  a_{40} =& \left(-\frac{7}{81}+\frac{5\pi^2}{216}\right) \,\delta_{l0}
   \nonumber \\ &
   + \left( \frac{119}{36}-\frac{\pi^2}{3}\right)\,\frac{j(j+1)-l(l+1)-3/4}{l(l+1)(2l+1)}
   (1-\delta_{l0})
  \,,\\
 & a_{50} = -0.229053\,\delta_{l0}\,,\\
 & a_{61} = \frac{a_{40}}2\,\delta_{l0}\,.
\end{align}
The higher-order remainder $G(\Za)$ inferred from our numerical results is plotted
in Fig.~\ref{fig:svpe}. For hydrogen, our results are: $G_{1s}(\alpha) =
0.93\,(6)$, $G_{2s}(\alpha) = 1.04\,(5)$, $G_{2p_{1/2}}(\alpha) = 0.02\,(4)$, and
$G_{2p_{3/2}}(\alpha) = 0.01\,(3)$. These values are consistent with the
$\Za$-expansion results for the normalized difference of the $2s$ and $1s$ states
and for the fine-structure difference \cite{jentschura:05:sese}:
$a_{60}(2s)-a_{60}(1s) = 0.109999$ and $a_{60}(2p_{3/2})-a_{60}(2p_{1/2}) =
-0.013435$.

For completeness, we specify also the result for the leading term of the $\Za$
expansion known for the S(VP)E correction beyond the free-loop approximation
\cite{pachucki:93:pra,eides:95:pra},
\begin{equation}\label{b4}
    \Delta E_{\rm SVPE,b} =
        m\,\left(\frac{\alpha}{\pi}\right)^2\, \frac{(\Za)^5}{n^3}\,
         \bigl[ -0.38615\,\delta_{l0}\bigr]\,.
\end{equation}


%
\section{Conclusion} \label{sec:concl}

In the present investigation, we performed calculations of the part of the
two-loop Lamb shift induced by the diagrams in Fig.~\ref{fig:2order}(d)-(k).
Numerical results were obtained for the $n=1$ and $n=2$ states and for the whole
region of the nuclear charge numbers $Z=1-100$. The diagrams (d)-(g) were
calculated rigorously to all orders in $\Za$, whereas for the diagrams (h)-(k),
the fermion loops were approximated by their leading $\Za$-expansion contribution.
An estimate was given for the higher-order terms thus omitted. The finite nuclear
size effect was accounted for in the evaluations of all diagrams except for the
diagram (k). The latter diagram was calculated with the point nuclear model; an
estimate of the finite nuclear size effect was supplied.

In the low-$Z$ region, our numerical results were employed for the identification
of the nonperturbative remainder $G(\Za)$, which incorporates all orders in the
$\Za$-expansion starting with $\alpha^2(\Za)^6$. For hydrogen, the net result for
the two-loop diagrams with closed fermion loops is
\begin{eqnarray}\label{c01}
    G_{1s}(\alpha) &=& -15.0(4)(2.2)\,, \\
    G_{2s}(\alpha) &=& -14.0(4)(2.2)\,, \\
    G_{2p_{1/2}}(\alpha) &=& -0.28(4) \,, \\
    G_{2p_{3/2}}(\alpha) &=& -0.05(3)  \,,
    \label{c01a}
\end{eqnarray}
where the first error quoted is the numerical uncertainty. The second error (if
given) is due to contributions beyond the free-loop approximation. We estimate
them for the $ns$ states as $1$ [in units of $G(\Za)$] for the VPVP correction and
as $2$ for the S(VP)E diagram (which arises as a typical coefficient of $0.2$
enhanced by the first power of logarithm). It is interesting to note that the
dominant part of the remainder term for the $ns$ states is due to the SEVP
correction [Fig.~\ref{fig:2order}(d)-(f)].

In order to get the complete results for the two-loop Lamb shift, one should
combine the numerical values obtained in the present work with the contribution
due to the two-loop self-energy [Fig.~\ref{fig:2order}(a)-(c)]. Its all-order
calculation was accomplished in our previous investigations, in
Refs.~\cite{yerokhin:03:prl, yerokhin:03:epjd,yerokhin:05:sese,yerokhin:05:jetp}
for the $1s$ state and $Z\geq10$ and in Ref.~\cite{yerokhin:06:prl} for the $n=2$
states and $Z\geq60$. Combined together, our calculations yield results for the
total two-loop QED correction, which improve the total theoretical accuracy of the
$1s$ Lamb shift \cite{yerokhin:03:prl} and that of the $2p_{1/2,3/2}-2s$
transition energy in heavy Li-like ions \cite{yerokhin:06:prl}.

Still, the project of the calculation of the two-loop Lamb shift is far from being
finished. There are several reasons for this. First, the results of the all-order
calculation of the two-loop self-energy correction for the $1s$ state in the
low-$Z$ region \cite{yerokhin:05:sese} do not agree well with the $\Za$-expansion
result of Ref.~\cite{pachucki:03:prl}. Second, the calculation
\cite{yerokhin:06:prl} for the $n=2$ states was performed in the high-$Z$ region
only. Third, a part of the two-loop diagrams with closed fermion loops is
presently calculated within the free-loop approximation only. Each of these points
represents a difficult problem and all of them should be solved before the
calculation of the Lamb shift to order $\alpha^2$ is completed.

Valuable discussions with K.~Pachucki and U.~D.~Jentschura are gratefully
acknowledged. V.A.Y. acknowledges support from RFBR (grant No.~06-02-04007) and
the foundation ``Dynasty.'' Laboratoire Kastler Brossel is Unit\'e Mixte de Recherche du CNRS n$^{\circ}$
8552, of the Physics Department of \'Ecole Normale Sup\'erieure and
Universit\'e Pierre et Marie Curie.

%


\end{document}